\documentclass[prl,twocolumn,showpacs,amsmath,amssymb]{revtex4}

\usepackage{graphicx}% Include figure files
\usepackage{dcolumn}% Align table columns on decimal point
\usepackage{bm}% bold math

\usepackage{color}

\begin{document}

%\pdfoutput=1
\newcommand{\bn}{{\bf n}}
\newcommand{\bp}{{\bf p}}   
\newcommand{\br}{{\bf r}}
\newcommand{\bk}{{\bf k}}
\newcommand{\bv}{{\bf v}}
\newcommand{\bj}{{\bf j}}
\newcommand{\wk}{\omega_{\bf k}}
\newcommand{\nk}{n_{\bf k}}
\newcommand{\eps}{\varepsilon}
\newcommand{\la}{\langle}
\newcommand{\ra}{\rangle}
\newcommand{\be}{\begin{eqnarray}}
\newcommand{\ee}{\end{eqnarray}}
\newcommand{\intl}{\int\limits_{-\infty}^{\infty}}
\newcommand{\dE}{\delta{\cal E}^{ext}}
\newcommand{\SE}{S_{\cal E}^{ext}}
\newcommand{\dsp}{\displaystyle}
\newcommand{\phit}{\varphi_{\tau}}
\newcommand{\p}{\varphi}
\newcommand{\cL}{{\cal L}}
\newcommand{\dphi}{\delta\varphi}
\newcommand{\dbj}{\delta{\bf j}}

\newcommand{\rred}[1]{{#1}}
\newcommand{\skp}[1]{{#1}}
\newcommand{\bblue}{}

\title{ Magnetic field induced non-Gaussian fluctuations in macroscopic %classical 
equilibrium systems}

\author{ K. E. Nagaev$^1$, O. S. Ayvazyan$^2$, N. Yu. Sergeeva$^2$, and M. B\"uttiker$^3$ }
\affiliation{$^1$Institute of Radioengineering and Electronics,  Mokhovaya 11-7, Moscow, 125009 Russia\\
             $^2$Moscow Institute of Physics and Technology, Dolgoprudny,  141700 Russia\\
             $^3$Department of Theoretical Physics, University of Geneva, CH-1211 Geneva 4, Switzerland}

\date{\today}

\begin{abstract}
We calculate the magnetic-field dependent nonlinear conductance and noise in a two-dimensional macroscopic inhomogeneous system. If the system does not possess a specific symmetry, the magnetic field induces a nonzero third cumulant of the current even at equilibrium. This cumulant is related to the first and second voltage derivatives of the spectral density and average current in the same way as for mesoscopic quantum-coherent systems, 
but these quantities may be much larger.
The system provides a robust test of a non-equilibrium fluctuation relation. 
\end{abstract}
\pacs{73.21.Hb, 73.23.-b, 73.50.Lw}

\maketitle

It is commonly believed that equilibrium fluctuations in macroscopic systems are Gaussian \cite{Landau}.
If their correlation length $l_c$ is much smaller than all the dimensions of the sample $L$, this is a direct
consequence of the central limit theorem, which says that the sum of a large number of random variables will have approximately normal distribution.
This is why higher cumulants of current are usually calculated and measured for mesoscopic systems, whose size is smaller than $l_c$. Typically, $l_c$ is the inelastic scattering length  and 
if $L<l_c$, the conductor behaves as a single quantum scatterer. 
In a macroscopic system, higher cumulants are usually much smaller than the second one.
However some fluctuations do not exponentially
decay with distance and therefore do not have a definite correlation length. For example, fluctuations of charge density in two-dimensional conductors are screened according to a power law. If they are nonlinearly coupled to the current, this may result in a non-Gaussian noise even in an equilibrium macroscopic system. Below we propose a model of such noise. Moreover, the nonzero equilibrium third cumulant of current in it obeys the fluctuation relations recently 
derived for mesoscopic systems.

The last several years saw an increase of interest to  fluctuation--dissipation relations that apply beyond the linear transport regime \cite{Tobiska05,Andrieux}. Based on the symmetry properties of the cumulant generating function in the full counting statistics of transferred charge \cite{Levitov93},  universal relations between cumulants of current of different order and their voltage derivatives were obtained. 

%Presently, it is not known exactly what sets the limits for this universality. It was
%of interest to check whether the universality resulted from microscopic reversibility,
%i.\  e.\ the invariance of the microscopic equations of evolution with respect to the
%sign of the time variable. The micro-reversibility should be broken by a magnetic field
%if one considers the nonlinear transport regime.  Indeed in mesoscopic physics, for
%quantum coherent systems, it has been shown theoretically and confirmed by experiment
%that the current-voltage characteristic is not even in magnetic field already to
% second order in voltage. However 

Recent studies \cite{Forster08,Saito08,Sanchez09} showed that the fluctuation relations hold even in a magnetic field that breaks the microscopic reversibility. In particular, they link the  contribution to the mean current proportional to magnetic field and quadratic in voltage $\Delta I \propto V^2H$ to a noise contribution proportional to the temperature, magnetic field, and voltage $\Delta S \propto THV$, and to the third cumulant of equilibrium current fluctuations $C_0$. Consequently, this cumulant is an odd function of the magnetic field. However the relations obtained in \cite{Forster08} and \cite{Saito08} differ. F{\"o}rster and B{\"u}ttiker \cite{Forster08} obtained that $C_0 =3T\,[\partial\Delta S/\partial V - T\,\partial^2\Delta I/\partial V^2]$, whereas Saito and Utsumi \cite{Saito08} obtained a stronger condition $C_0 = 2T\,\partial\Delta S/\partial V = 6T^2\,\partial^2\Delta I/\partial V^2$ valid for a more restricted class of systems. The weaker condition permits magnetic field asymmetric conductance and noise even if the third cumulant of the equilibrium noise vanishes. 

The nonlinear contributions to the average current $\Delta I \propto V^2H$ were calculated for a number of mesoscopic systems. Sanchez and Polianski and one of the authors \cite{Sanchez04,Polianski06} investigated a quantum Hall bar with an
antidot and a chaotic cavity connected to quantum point contacts. Spivak and Zuyzin \cite{Spivak04} calculated this contribution for a mesoscopic diffusive system, whereas Andreev and Glazman \cite{Andreev06} studied it for a semiclassical ballistic contact with an asymmetric obstacle. This contribution was observed in several experiments \cite{Zumbuhl06,Leturcq06,Angers07,Brandenstein09}. A qualitative relation between  the magnetic field asymmetric current $\Delta I$ and the voltage derivative of the noise  $\Delta S$ was established in a pioneering experiment  for an Aharonov--Bohm interferometer \cite{Nakamura10}.

\begin{figure}[t]
\includegraphics[width=7.5cm]{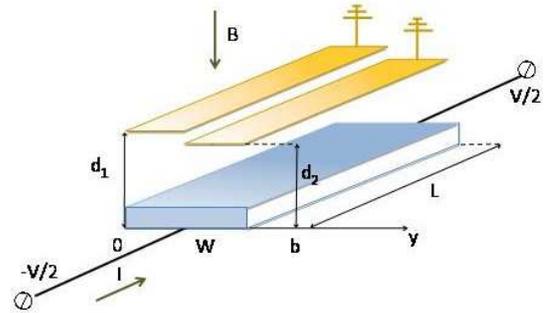}
 \caption{\label{fig1} A sketch of a diffusive conductor with an asymmetric gate. Its I-V-characteristics, noise and the third cumulant of equilibrium noise are magnetic-field asymmetric. Gap heights $d_1$ and $d_2$ correspond to $\alpha_1$ 
and $\alpha_2$ in Eq. (\ref{I_series}). }
\end{figure}

The fluctuation relations in a magnetic field \cite{Forster08,Saito08} were derived for  mesoscopic systems assuming conservation of the electron energy. 
Therefore it is of interest to test them for a macroscopic system with a strong energy dissipation by explicitly calculating the relevant quantities. 
Here we consider a system that can be made arbitrarily large and yet exhibits non-Gaussian noise due to long-range charge fluctuations. We use it to test the nonlinear fluctuation--dissipation relations in a magnetic field. 
Because of the macroscopic nature of the system, the quantities in hand may reach much higher values than in mesoscopic conductors and are more easy to measure. 

%and present the calculation of the non-linear current-voltage characteristic, the %voltage dependent noise and the third cumulant at  equilibrium. 
%Interestingly the effects we find are macroscopic. In particular the third cumulant of
% equilibrium noise is non-vanishing and macroscopic. 
%That is the main difference to the theories and  the experiments discussed above. The
% robustness of our effect will bring a test of the first non-trivial fluctuation
% relation within experimental reach. 

\rred{{\it Model.} To perform our test, we need a macroscopic conductor that possesses a nonlinear conductivity and lacks
symmetry in the direction transverse to the current.}
The system of our choice (shown in Fig. 1) is a conducting slab with a thickness much smaller than the charge screening length and two perfect electrodes. The slab is threaded by a magnetic field normal to its surface, and a grounded gate is mounted on top of it. The narrow gap of height $d_g$ between the slab and the gate and the 2D conductivity of the slab $\sigma_0$ are spatially nonuniform. We also assume a diffusive transport and a strong energy relaxation so that the local distribution of electrons is in equilibrium and the local conductivity is determined solely by the electron sheet density $n_s$. 
The conductor and the gate form together a plane capacitor so that the electron density in the slab is  related to the local potential $\varphi(\br)$ by 
$
 e(n_s - n_0) = \eps_d\varphi/4\pi d_g,
$
where $n_0$ is the equilibrium electron density at $\varphi=0$ and $\eps_d$ is the dielectric constant of the insulator
in the gap. Therefore the  local conductivity depends  on the local potential, i.~e.
\be
 \sigma(\br)=\sigma_0(\br)\,[1 + \alpha(\br)\,\varphi(\br)],\qquad
\nonumber\\
 \sigma_0 = \frac{e^2 n_0(\br)\tau_{tr}}{m},
 \quad 
 \alpha(\br)=\frac{\eps_d}{4\pi e n_0(\br)\,d_g(\br)},
 \label{sigma_vs_phi}
\ee
where $m$ is the electron mass and $\tau_{tr}$ is its coordinate- and energy-independent scattering time. The sensitivity of $ \sigma(\br)$ to the potential  leads to a current which is  nonlinear in the applied voltage. 
If a current flows from one electrode to another in a magnetic field, it induces a Hall voltage across the conductor. If $\alpha(\br)$ is asymmetric in the direction transverse to the current, the different directions of current or magnetic field will result in different changes of conductivity and hence the antisymmetric component $\Delta I \propto V^2H$ should appear in the current. 

%\begin{figure}[t]
% \includegraphics[width=8.5cm]{fig2}
% \caption{\label{fig2} The spatial distribution of the Hall correction to the electric
% potential $\p_{\beta}$
% in a magnetic field for a rectangular conductor. The antisymmetric contribution to the % current (\ref{I_series})
% is obtained by averaging it with a spatially nonuniform coefficient $\alpha$}
%\end{figure}

In a magnetic field $H$, the local current density is 
\be
 \bj = -\sigma_0\,(1 + \hat\beta)(1 + \alpha\p)\,\nabla\p,
 \quad
 \hat\beta = 
 \left(
  \begin{array}{rr}
    0 & \beta\\ -\beta & 0 \\
  \end{array}
 \right),
 \label{j-general}
\ee
where the matrix $\hat\beta$
acts on the  vector $\nabla\p$ and $\beta =eH/(mc)\,\tau_{tr}$ is proportional to $H$.
The current density satisfies 
the continuity equation
$
 \nabla\cdot\bj =0\,.
$
The boundary conditions at the left and right electrodes are $\left.\p(\br)\right|_L = V_L$ and 
$\left.\p(\br)\right|_R = V_R$, where $V_R - V_L = V$. The rest of the conductor boundary is impenetrable to the current. 
Without loss of generality, we assume that $\sigma_0$ smoothly turns to zero at the insulating 
boundary whereas $\bn\cdot\nabla\p$ remains finite there.

\rred{\it Average current.}
We solve the continuity equation by expanding $\p$
and $\bj$ in $\alpha$ and $\beta$ , 
\be
 \p = \p_0 + \p_{\alpha} + \p_{\beta} + \p_{\alpha\beta},
 \quad
 \bj = \bj_0 + \bj_{\alpha} + \bj_{\beta} + \bj_{\alpha\beta}.
 \label{j-series}
\ee
%
%This leads to the equations
%
%
%\be
% {\cal L}\p_0 \equiv \nabla(\sigma_0\nabla\p_0) =0,
% \nonumber\\
% {\cal L}\p_{\alpha} = -\nabla(\alpha\p_0)\,\sigma_0\,(\nabla\p_0),
% \nonumber\\
% {\cal L}\p_{\beta} = -(\nabla\sigma_0)\,\hat\beta\,(\nabla\p_0),
% \label{phi_a-b}
%\ee
%
%where $\p_0$ satisfies the same boundary conditions as $\p$ and the terms of higher order
%in $\alpha$ and $\beta$ obey zero boundary conditions. 
%
In our derivation, we use an approach 
similar to that of \cite{Sukhorukov98}. The correction to the total current flowing through the 
conductor is conveniently expressed in terms of the characteristic potential  $\psi$ of the left contact that obeys
\rred{ an equation $\cL\psi \equiv \nabla\cdot(\sigma_0\nabla\psi)=0$} with boundary conditions $\left.\psi\right|_L = 1$ and 
$\left.\psi\right|_R = 0$ \cite{Buttiker93} so that $\p_0 = V_L\,\psi + V_R\,(1-\psi)$. \bblue{In the absence of a magnetic field, 
$\psi$ allows one to map the conductor on a purely one-dimensional system, {where it} plays the role of {the}  longitudinal
coordinate $x$.} For vanishing $\alpha$ and magnetic field, the current through the conductor 
can be written  in a form
%
%as an integral along the boundary $\Gamma$ of the conductor
%
%\be
% I_0 = \int_{\Gamma} dl\,\psi(l)\,{\bf n}\bj_0,
% \quad
% \bj_0 = -\sigma_0\nabla\p_0.
% \label{I0-1}
%\ee
%
% Transforming this contour integral into an integral over the 
%area of the conductor 
%making use of the condition $\nabla\bj_0=0$ and
% the relation $\nabla\p_0 = (V_L - V_R)\,\nabla\psi$, we obtain
%
\be
 I_0%=\int d^2r\,(\nabla\psi)\,\bj_0
 = %(V_R - V_L) 
 {V}\int d^2 r\,\sigma_0\,(\nabla\psi)^2.
 \label{I0-2}
\ee
%
%In a similar way, we 
We obtain the first order nonlinear correction in $\alpha$ to the current by
separating the component $\bj_{\alpha}$ in Eq. (\ref{j-general})
which takes the form of an integral
\be
 I_{\alpha} = %(V_R - V_L) 
 {V}\int d^2r\,\alpha\,\sigma_0\,\p_0\,(\nabla\psi)^2.
 \label{I_ab-1}
\ee
This quantity is proportional to the nonlinear correction to the local conductivity averaged over the 
conductor with a weigh factor $(\nabla\psi)^2$. It vanishes for a symmetric conductor if $V_L=-V_R$. 
The quadratic nonlinear conductance and the rectification effect were recently discussed in detail in 
\cite{Polianski07}.

The antisymmetric contribution to the current $I_{\alpha\beta}$ can be calculated by isolating the 
corresponding component in Eq. (\ref{j-general}).
%The potentials $\p_{\alpha}$ and $\p_{\beta}$ are obtained from Eq. (\ref{phi_a-b}) 
%by inverting the differential operator $\cL$. We find that 
 \rred{We} can present
$\Delta I\equiv I_{\alpha\beta}$ in \rred{the} form
%equals
%
\be
 \Delta I =
 {V^2}\!\int d^2r\,
 \alpha\,\sigma_0\,(\nabla\psi)^2
 \cL^{-1}\left[(\nabla\sigma_0)\cdot\hat\beta\,\nabla\psi\right],
 \label{I_ab-4}
\ee
where $\cL^{-1}$ is the inverse of differential operator ${\cal L}$ \skp
taken with zero boundary conditions.
This component of the current 
results from the Hall voltage 
and the related correction to the conductivity.
\bblue{Note that in a magnetic field, the problem cannot be mapped onto a one-dimensional one. Therefore
the operator $\cL^{-1}$ cannot be eliminated from the right-hand side of Eq. (\ref{I_ab-4}) in contrast 
to Eq. (\ref{I0-2}) for $\delta I_{\alpha}$.} 

\rred{\it Noise.} To find the spectral density of \bblue{electric} noise, i. e. the second cumulant of the current, we use
the Langevin approach \cite{Kogan96}. To this end, we linearize Eq. (\ref{j-general}) with respect 
to small fluctuations $\delta \p$ and $\delta\bj$ and add a random extraneous current $\delta\bj^{ext}$ 
to its right-hand side so that
\be
 \delta\bj = -\sigma_0(1 + \hat\beta)[\nabla\delta\p + \alpha\nabla(\p\,\delta\p)] +   
 \delta\bj^{ext}.
 \label{dj-basic}
\ee
 We are interested in the low-frequency fluctuations for which  the 
continuity equation $\nabla\cdot\delta\bj=0$ holds. The fluctuations also obey the boundary conditions $\delta\p=0$ at the electrodes and $\bn\,\delta\bj=0$ at the rest of the boundary. The continuity equation results in a diffusion-type
equation in $\delta\p$. Using the formal solution of this equation and Eq. (\ref{dj-basic}), we express the
fluctuation $\delta I$ in terms of $\delta\bj^{ext}$. 
%and then integrate it over the interface with one of the electrodes to obtain the fluctuation of total current $\delta I$.
Then the product of two different
realizations $\delta I$ has to be averaged using the correlation function of extraneous currents.
We assume that because of a strong energy 
relaxation the local distribution of electrons is equilibrium for a given electric potential and therefore this correlation function  is given by
\be
  \la\delta j_{x,y}^{ext}(\br_1)\,\delta j_{x,y}^{ext}(\br_2)\ra
 %=\la\delta j_y(\br_1)\,\delta j_y(\br_2)\ra
 =2T\,\delta(\br_1 - \br_2)\,
 \nonumber\\
 \times[1 + \alpha\,\p(\br_1)]\,\sigma_0(\br_1).
 %\nonumber\\
 %\la\delta j_x^{ext}(\br_1)\,\delta j_y^{ext}(\br_2)\ra 
 %\rred{\propto \sigma_{xy}(H) + \sigma_{yx}(H)}= 0.
 \label{extraneous}
\ee
Extraneous currents $\delta j_x^{ext}$ and $\delta j_y^{ext}$ are uncorrelated because their correlation
function is proportional \cite{Landau} to $\sigma_{xy}(H) + \sigma_{yx}(H)= (\sigma_0\hat\beta)_{12}
+ (\sigma_0\hat\beta)_{21}=0$.

We solve Eq. (\ref{dj-basic}) by expanding $\delta\p$ and $\delta\bj$ in $\alpha$ and $\beta$ similarly
to Eqs. (\ref{j-series}). %While solving the Langevin equation for the fluctuations, we consider $\delta\bj^{ext}$ 
%as zero-order in $\alpha$ and $\beta$. After two formal expressions for $\delta I$ have been multiplied and 
%averaged, the correlation function (\ref{extraneous}) should be also expanded in these parameters to the required
%order. 
In the zero approximation, we obtain a formal expression for the current fluctuation
$
 \delta I_0 %= \int\limits_{\Gamma} dl\,\psi(l)\,{\bf n}\,\delta\bj_0
 = \int d^2r\,(\nabla\psi)\cdot\delta\bj^{ext},
 %\label{dI0}
$
which leads to the spectral density of current of the form

\be
 S_0 = 2T\int d^2r\,\sigma_0\,(\nabla\psi)^2
 \equiv 2T\,\partial I_0/\partial V
 \label{dI^2_0}
\ee
that satisfies the usual Nyquist theorem. 

The correction to the spectral density proportional to voltage equals
%is calculated using the formal expression
%
%$\delta I_{\alpha} =-\int d^2r\,\sigma_0(\nabla\psi)\,\alpha\,\nabla(\p_0\,\dphi_0)$, 
%
%where
%
%$\dphi_0 = \cL^{-1}(\nabla\dbj^{ext})$. Making use of the symmetry of $\cL^{-1}$ and 
%integrating by parts gives
%is
%
%\be
 %\delta I_{\alpha} =-\int d^2r\,\nabla\!
 %\left\{
  %\cL^{-1}[\sigma_0\,(\nabla\psi)(\nabla\alpha)\,\p_0]
 %\right\} \dbj^{ext}.
 %\label{dI_a-4}
%\ee
%
%By multiplying $\delta I_{\alpha}$ and $\delta I_0$ and using the correlation function (\ref{extraneous}) to the
%zero order in $\alpha$
%, integrating the result by parts and making use of the equation $\cL\psi=0$, 
%It is easy 
%to show that the correlation function $\la\delta I_{\alpha}\,\delta I_0\ra$
%is zero in linear approximation in $\alpha$. Hence the only nonvanishing contribution to %$S_{\alpha}$ 
%arises from the
%term $2T\alpha\sigma_0\p_0$ in the correlator of extraneous currents (\ref{extraneous}) %and
%
\be
 S_{\alpha}
 =2T \int d^2r\,\alpha\sigma_0\p_0\,(\nabla\psi)^2
 \equiv 2T\,\partial I_{\alpha}/\partial V.
 \label{dI^2_a-2}
\ee
The second part of this equation is exactly the relation established in Refs. \cite{Forster08,Saito08} for the
quantities symmetrized with respect to the magnetic field.

The correction to the noise $S_{\alpha\beta}$ proportional to both voltage and magnetic field is determined
by the components of the current fluctuation $\delta I_0$, $\delta I_{\alpha}$, and $\delta I_{\beta}$. 
%The latter is given by
%
%\be
 %\delta I_{\beta} = \int d^2r\,
 %\nabla\!\left\{\cL^{-1}
 % \left[(\nabla\sigma_0)\,\hat\beta\,\nabla\psi\right]
 %\right\}
 %\dbj^{ext}.
 %\label{dI_b-2}
%\ee
%
%The component $\delta I_{\alpha\beta}$  {may be presented in} the  form
%
%$\delta I_{\alpha\beta} = \int d^2r\,\nabla\!\left\{\cL^{-1}[\ldots]\right\}$. 
%
%Using the same reasoning as in the case of $S_{\alpha}$, it 
%It is easily obtained that the correlator $\la\delta I_{\alpha\beta}\,\delta I_0\ra$ is zero to the required order, 
%hence $\delta I_{\alpha\beta}$ does not 
%contribute to $S_{\alpha\beta}$. 
Long but simple calculations lead to the expression for $\Delta S \equiv S_{\alpha\beta}$
of a form
\be
 \Delta S 
 %\equiv S_{\alpha\beta}
 =6T\,%(V_R - V_L)
 {V} \int d^2r\,\alpha\sigma_0\,(\nabla\psi)^2\,
 %\nonumber\\ \times
 \cL^{-1}\left[(\nabla\sigma_0)\cdot\hat\beta\,\nabla\psi\right],
 \label{dI^2_ab-2}
\ee
hence the antisymmetric contribution to the noise satisfies the condition
$\partial\Delta S/\partial V = 3T\,\partial^2\Delta I/\partial V^2$ \cite{Saito08}. The presence of 
this contribution to the noise suggests that in a nonzero magnetic field, the minimum in the voltage dependence of noise \cite{Forster09} is shifted away from $V=0$.

\rred{\it Third cumulant.} To calculate the equilibrium third cumulant of current, we use the semiclassical 
cascade approach
\cite{Nagaev02,Pilgram03,Jordan04}. We assume that the conductor is diffusive and therefore the local
random Langevin currents have a Gaussian distribution with zero third cumulant \cite{Nagaev02}. 
However these random currents induce fluctuations of the potential $\delta\p$, which may affect the correlator
of Langevin currents at a different point by changing the local conductivity (\ref{sigma_vs_phi}) that
enters into Eq. (\ref{extraneous}). This results in an irreducible correlation between three 
observable currents. The third cumulant of current is 
\be
 C = 3\int d^2r\,\frac{\delta S}{\dphi(\br)}\,
 \la\dphi(\br)\,\delta I\ra,
 \label{C-1}
\ee
where ${\delta S}/{\dphi(\br)}$ is the functional derivative of the spectral density of the current $S$ with
respect to the local potential $\p$ at point $\br$. As this derivative is proportional to $\alpha$, 
the correlator $\la\delta\p(\br)\,\delta I\ra$ should be calculated to the zero approximation in this
parameter. 
%To obtain it, one has to calculate averages of a type
%
%\be
% \left\la{\cal A}[\nabla\dbj^{ext}(\br)] \int d^2r'\,{\bf B}(\br')\,\dbj^{ext}(\br')\right\ra
% \nonumber\\
% \equiv 2T{\cal A}\{\nabla[\sigma_0{\bf B}(\br)]\},
% \label{<AB>}
%\ee
%
%where $\cal A$ is an arbitrary linear operator and $\bf B$ is an arbitrary vector function. Applying this
%formula, it is easily obtained that in 
In the required approximation, the sought-for correlator equals
\be
 \la\dphi(\br)\,\delta I\ra
 =2T\cL^{-1}\!\left[(\nabla\sigma_0)\cdot\hat\beta\,\nabla\psi\right]
 \label{<dphidI>_b}
\ee
and vanishes in zero magnetic field. Thus ${\delta S}/{\dphi(\br)}$  is needed only to the zero order in $\beta$ and therefore
\be
 C_0 = 12\,T^2\int d^2r\,\alpha\sigma_0\,(\nabla\psi)^2\,
 \cL^{-1}\!\left[(\nabla\sigma_0)\cdot\hat\beta\,\nabla\psi\right].
 \label{C-2}
\ee
$ C_0 $ is proportional to $T^2$ and vanishes in zero magnetic field.
Like the asymmetric part of $S$, it is proportional to the asymmetric part
of the nonlinear conductance and satisfies
{the relation} 
$C_0 = 6T^2\,\partial^2\Delta I/\partial V^2$ %in accordance with 
 \cite{Saito08}.
In the quantum-coherent approach \cite{Forster09}, the $T^2$ dependence of the 
third cumulant was explained 
%by relating this quantity 
with the energy dependence 
of electron transmission through the conductor. In semiclassics, it arises quite naturally
because Eq. (\ref{C-2}) actually involves a product of two correlators of thermal noise 
(\ref{extraneous}). At equilibrium and in the absence of magnetic field the local potential  fluctuations do not contribute to the fluctuation of total current,
{and therefore $\delta\p$ and $\delta I$} %they 
are uncorrelated. 
It is the magnetic field that introduces a correlation between them as the fluctuation
$\delta I$ immediately leads to a fluctuation of the Hall voltage in the transverse direction. To 
exhibit a nonvanishing third cumulant of current, the system must possess a certain asymmetry so that
the fluctuation of the Hall voltage is not averaged out upon integration over its area.

\begin{figure}[t]
\includegraphics[width=8.5cm]{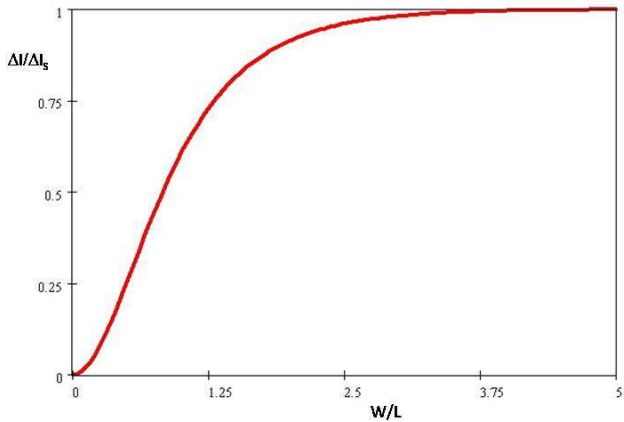}
 \caption{\label{fig3} The ratio of the term %asymmetric contribution to the current 
 $\Delta I$ for a rectangular conductor to its saturation value 
 $\Delta I_S = (1/12)\,\beta\,\sigma_0\,(\alpha_1 - \alpha_2)\,V^2$ as a function of   
 its width-to-length ratio.}
\end{figure}

\rred{\it Example.} Consider a specific example of a rectangular conductor with $\sigma_0=$const and 
dimensions $0<x<L$ and $0<y<W$ with the
current flowing in the $x$ direction. Assume that the gate above the conductor is split in the direction
transverse to the current so that $\alpha(y) = \alpha_1\Theta(W/2-y) + \alpha_2\Theta(y-W/2)$, where
$\Theta(y)$ is the Heaviside step function. Calculations by means of Eq. (\ref{I_ab-4}) give
\be
 \Delta I =
 \frac{4}{\pi^4}\,\beta\,\sigma_0\,(\alpha_1 - \alpha_2)\,V^2
\nonumber\\
 \times
 \sum_{n=0}^{\infty} \frac{1- (-1)^n}{n^4}
 \left[ 1 - \frac{1}{\cosh(n\pi W/2L)} \right].
 \label{I_series}
\ee
For long and narrow conductors with $W \ll L$ this expression reduces to $(1/8)\,\beta\,\sigma_0\,(\alpha_1 - \alpha_2)\,V^2\,(W/L)^2$. 
In the opposite limit of short and wide contacts $L \ll W$ it saturates as a function of 
$W/L$ and tends to $(1/12)\,\beta\,\sigma_0\,(\alpha_1 - \alpha_2)\,V^2$. The $\Delta I(W/L)$ curve is shown in 
Fig. \ref{fig3}. For a square conductor, the $\Delta I/I_0$ ratio does not depend on its size and appears to be much larger than for a mesoscopic system.
%The ratio of $\Delta I$ to that obtained in \cite{Andreev06} is of the order
%of $l^2_{tr}\,a_0/d^2\,d_g$, where $l_{tr}$ is the electron mean free path, 
%$a_0 =\eps_d\,\hbar^2/m\,e^2$ is the Bohr radius of an electron in the conductor, and %$d$ the size of a ballistic dot in \cite{Andreev06}. As $l_{tr}/d > 1$, this ratio may
%be of the order of 1. 
For a quantum-coherent mesoscopic system of size $d$ one can use an estimate 
$\Delta I_Q \sim e (eH/mc)(eV\tau_d/h)^2$, where $\tau_d$ is the dwell time of an electron in the system and $h$ is the Planck's constant \cite{Sanchez04,Spivak04}. As the maximum  $H$ and $V$ in this equation can be evaluated from the conditions 
$Hd^2 \sim \Phi_0 \equiv hc/e$ and $eV\tau_d \sim h$, one obtains the maximum value of the asymmetric current of the order of
$\Delta I_Q \sim e(\eps_F/h)(\lambda_F/d)^2$, where $\eps_F$ is the Fermi energy and $\lambda_F$ is the Fermi wavelength. An estimate of Eq. (\ref{I_series}) for a degenerate two-dimensional electron gas made on the assumption that $\alpha V \sim 1$ and $\beta \sim 1$ gives us 
$\Delta I_M \sim e(\eps_F/h) (l_{tr} d_g/a_B\lambda_F)$, 
where $l_{tr} = v_F\tau_{tr}$ and $a_B = \eps_d \hbar^2/me^2$ is the Bohr radius of an electron in the conductor. The ratio of these quantities 
$\Delta I_Q/\Delta I_M \sim \lambda_F^3a_B/d^2 d_g l_{tr}$ is much smaller than unity.

\rred{\it Discussion.}
Equations (\ref{dI^2_ab-2}) and (\ref{C-2}) suggest that the system exhibits a component of noise $\Delta S\propto VH$ and an equilibrium third cumulant of current $C_0 \sim T^2\beta\sigma_0(\alpha_1 - \alpha_2)$. The ratio of $C_0$
to the conductance is independent of the size of the square conductor precisely as for a mesoscopic system. As the third cumulant is insensitive to inelastic scattering, it can be made much larger than in mesoscopic systems by increasing temperature.

%The effect cannot be made arbitrarily large by increasing one or both of the dimensions
% of the conductor and remains ``mesoscoscopic'' in that sense.

\rred{In summary,} 
we have made predictions for the non-linear conductance, the voltage dependence of  noise and the third cumulant  in a macroscopic system with a strong energy dissipation. Long-range fluctuations of charge along with nonlinearity, macroscopic inhomogeneity  and the magnetic field lead in particular to a nonzero third cumulant of current at equilibrium. This cumulant is proportional to the magnetic-field-asymmetric nonlinear conductance and voltage-dependent noise
%The proportionality coefficients are the same as for quantum-coherent transport and satisfy a non-equilibrium %fluctuation relation. 
and satisfies the stronger form of the two fluctuation relations derived for quantum-coherent transport. {\it This suggests 
that quantum coherence or energy conservation are not necessary for these relations to hold and they may be considered
as universal.} 
%The macroscopic nature should facilitate measurement of all elements of the relation
% including the third cumulant.
The macroscopic size of the third cumulant makes its measurement possible
and thus could provide an experimental confirmation of the first non-trivial
fluctuation relation.

{\it Acknowledgments.}
This work is supported by the Swiss NSF, the center for excellence MaNEP and the European ITN NanoCTM.

\end{document}